\documentclass[prl,10pt,twocolumn,showpacs]{revtex4}
\usepackage{dcolumn,bm,amsmath,graphicx}
\begin{document}
\title{Nonlinear Noise Reduction Scheme Based on Information Flow}
\author{Seung Ki Baek}
\email[e-mail: ]{garuda@kaist.ac.kr}
\affiliation{Department of Physics, Korea Advanced Institute of Science and Technology, Daejeon 305-701, Republic of Korea}
\begin{abstract}
We present a measurement noise reduction scheme based on information flow of a chaotic system. This scheme operates on conditions of chaoticity and well-defined noise level, not depending on other detailed characteristics of noise. Starting with a simple map and full knowledge of dynamics, we extend the basic idea to general form applicable to higher dimensional systems. Reducing noise in Lorenz system is demonstrated as an example. Inferring dynamics without {\em a priori} knowledge is then discussed by proposing an indicator which measures predictability.
\end{abstract}
\maketitle
It has been of great importance in communication and experimental research how to filter off noisy parts from the signal. As the broad-band spectrum of signals from nonlinear chaotic systems usually makes traditional linear filters unfeasible, many researchers have studied noise reduction methods applicable to nonlinear systems \cite{Kost1988, Kost1989, Hammel1990, Farmer1991, Davies1994, Schreiber1993, Kost1993, Davies1998, Brocker2001}. It is widely known that there are two kinds of noise : {\em measurement noise} means corruption of data in observation process without interfering dynamics itself, while {\em dynamical noise} denotes the perturbation of the system coupled to dynamics, occurring at each time step. The noise reduction problem is quite different for each case and we treat measurement noise in this paper. There exists a {\em true} orbit $\{Y_k\}_{k=1}^{N}$ satisfying certain dynamics $Y_{k+1} = M(Y_k)$ for $1 \le k \le N-1$, while one observes only a {\em noisy} orbit $\{X_k\}_{k=1}^{N}$ given by $X_k = Y_k + \eta_k$ for small $|\eta_k|<\delta$, where $\eta_k$ and $\delta$ denotes {\em noise} and {\em noise level}, respectively. We would like to obtain a less noisy orbit $\{X'_k\}_{k=1}^{N}$, and most approaches take this problem through minimizing a target function with constraints, such as
\begin{equation}
\label{Lagrangian}
S = \sum_{k=1}^{N}|X'_k - X_k|^2 + \sum_{k=1}^{N-1}\{M(X'_k)-X'_{k+1}\} \lambda_k,
\end{equation}
where $\lambda_k$ is a Lagrangian multiplier \cite{Farmer1991}. Minimizing $S$ corresponds to maximizing {\em likelihood} function $\bar{P}$ within a time interval $[t-\alpha, t+\beta]$ :
\begin{eqnarray}
\label{likelihood}
&\bar{P}(&M^\alpha(X_{t-\alpha}),\ldots, M^{-\beta} (X_{t+\beta}))\nonumber\\
&\propto& \prod_{j=\alpha}^{j=-\beta} \exp \left( -\frac{1}{2 \sigma^2} \left| \frac{M^j (X_{t-j})-Y_t}{dM^j (Y_{t-j})} \right|^2 \right)
\end{eqnarray}
where $dM$ is the derivative of $M$, under the assumption that the sequence $\{\eta_k\}$ is independently Gaussian distributed with standard deviation $\sigma$. Those probability distributions of position at different times are transported to a particular time, distorted by chaotic dynamics $M$, and the true data point is restricted to their intersection. Thus maximum of joint probability function $\bar{P}$ estimates the position of true data point at that time. We shall discuss how this calculation is simplified if we consider information aspects as in communication area.

Studies on communication using chaos \cite{Hayes, Bollt, Cuomo} has been carried out from the understanding of chaos control\cite{OGY, Kantz} and chaos synchronization \cite{Pecora}. The main issues in this field are how to encode information using chaotic signal with dynamics already known to both of transmitter and receiver, and how to build a system persistent from noise occurring in communication channel, which corresponds to measurement noise. Rosa et al. \cite{Rosa} illustrated a filtering method using $2x \bmod 1$ map. This method, which will be called {\em Rosa's method}, is described as following : one picks a point $(X_t, X_{t+1})$ and executes backward iteration on $X_{t+1}$ resulting in two preimages $\hat{X}_t^{Left}$ and $\hat{X}_t^{Right}$, one of which closest to $X_t$ is selected as a filtered point of time $t$. This filter shrinks noise by a factor of two (i.e. Lyapunov exponent of the map) at each iteration. Andreyev et al. \cite{Andreyev} investigated information aspects and applications of Rosa's method. They, however, only treated basically 1-dimensional maps since they had to operate inverse mapping directly.

Maximum likelihood method and Rosa's method are actually identical although the former originates from the topological distortion \cite{Wolf} and the latter from information property. In a viewpoint of information theory \cite{Shannon, Brillouin}, a chaotic system itself is interpreted as an active processor of information \cite{Ott}. Supposing we have a measuring tool with finitely limited resolution, stretching process reveals the initial state impossible to identify with the tool at that time more precisely. If only stretching process exists, the occupied areas in state space, i.e. the energy of the system diverges to infinity as the precision increases infinitely, as Brillouin claimed in Ref.\ \cite{Brillouin}. Folding process prevents this divergence, removing some stored information  inevitably, so we cannot discriminate every detail of the past merely by observing the present state. Topological distortion, therefore, induces the flow of information bits and successive recording of this flow determines more precise knowledge of the state in chaotic systems. Information flow is a general property of chaos and, for example, all hyperbolic chaotic systems are already proven to have constant positive information rates by Schittenkopf and Deco \cite{Schittenkopf}. This idea forms the basis of our scheme which connects two previous methods. First, we begin with fully known dynamics, just as in communication, and discuss later how to deal with given data without {\em a priori} knowledge.

Following Rosa et al., we start with the case of $2x \bmod 1$ map as the simplest example of stretch-fold mechanism and also of our scheme. Employing binary representation in describing states, each iteration simply shifts the decimal point one space to the right. Let us assume that we introduce noise with such a level that we can guarantee only the first effective number. If the initial state is observed to be $0.a_0 x x  \ldots$ and the first and second iteration give $0.a_1 x x \ldots$ and $0.a_2 x x \ldots$, respectively, noting that digits marked by $x$ may be spurious, we can say that the initial state is in fact $0.a_0 a_1 a_2 \cdots$, effectively reducing the noise on the initial state.

The above example involves two conditions : the noise level $\delta$ is known and the dynamics is chaotic. In such cases, we ignore the spoiled parts and that converts an observed point to a set of candidate points leading to degeneracy (e.g. all the points whose first digit is $a_0$). Then we clarify what it should be by receiving information from other unspoiled parts of data. Roughly speaking, proper temporal extension can compensate spatial ambiguity \cite{Pethel}. If a data point $X_t$ is observed, the real value $Y_t$ should lie within a finite neighborhood $I(X_t)$, whose size comes from the noise level $\delta$. The next real value $Y_{t+1}$, evolving from $Y_t$ deterministically, also belongs to $I(X_{t+1})$ while it does not hold for every point $p_t \in I(X_t)$ and its successor, $p_{t+1}$. Noting that the inverse mapping $M^{-1}$ operates on a set of points, not on a single point where the inverse map cannot be defined, we find the $n$-th order refinement,
\begin{equation}
\label{refinement}
I(X_t)^{new}_{(n)} = \bigcap_{i=0}^{n} M^{-i}\left\{I(X_{t+i})\right\}.
\end{equation}
In terms of the previous example, $M^{-i}\left\{I(X_{i})\right\}$ with $t=0$ means the set of binary numbers whose $i$-th digit is $a_i$. As the $n$-th order refinement requires $n+1$ successive measurements, it is obvious that the diameter of a remaining set never increases so that this algorithm is convergent :
\begin{equation}
\label{convergence}
0 < \left| I(X_t)^{new}_{(n)} \right| \leq \left| I(X_t)^{new}_{(n-1)} \right|.
\end{equation}
Equation (\ref{refinement}) shares similarity with (\ref{likelihood}) of maximum likelihood method, while Gaussian assumption is turned out to be unnecessary in our scheme. Once $\delta$ is defined, other details of noise are irrelevant. It is also worth noting that (\ref{refinement}) formalizes the basic philosophy of Rosa's method. The difficulty in application of it is remedied by rewriting (\ref{refinement}) as following :
\begin{equation}
\label{rewriting refinement}
M^n \left\{I(X_t)^{new}_{(n)}\right\} \subset \bigcap_{i=0}^{n} M^{i}\left\{I(X_{t+n-i})\right\}
\end{equation}
and this allows one to avoid calculating inverse mapping, hardly possible in high dimensional systems. We deduce that if a point does not belong to the set of the right-hand side of (\ref{rewriting refinement}), it cannot lie in the set of the left-hand side. Then what has to be done is only selecting points within $I(X_t)$ which satisfy the right-hand side after $n$ times of mapping. Henceforth, we iterate all nearby grid points around the observed data which approximate $I(X_t)$ in a discrete manner, and reject false ones getting outside the next expected intervals, $I(X_{t+1})$. We repeat the same procedure only on the surviving points until the number of remaining ones are less than a certain threshold, i.e. $\left| I(X_t)^{new}_{(m)}\right| < R_{th}$. $X_t$ is then corrected to $X'_k = \left<I(X_t)^{new}_{(m)}\right>$, the average of those remaining points. The number of steps $m$, required to reach this threshold $R_{th}$, measures the performance of noise reduction and we define this quantity as {\em abrasion time}. Since each point has its $m$, we obtain another sequence of abrasion time $\{m_k\}_{k=1}^{N}$ after refinement. A system with short $m$ is so sensitive that wrong guesses are easily rejected, and thus it is easy to clean noise. Later in inferring dynamics without knowledge of it, we use this concept in a different context, that is, fast abrasion implies large deviation from the true dynamics.

Figure 1 demonstrates the result of this scheme for Lorenz system :
\begin{equation}
\label{Lorenz}
\left\{\begin{array}{rcl}
\dot{x} &=& \sigma (y-x)\\
\dot{y} &=& rx - y - xz\\
\dot{z} &=& xy - bz
\end{array}\right.
\end{equation}
\begin{figure}
\includegraphics[angle=0, width=0.4\textwidth]{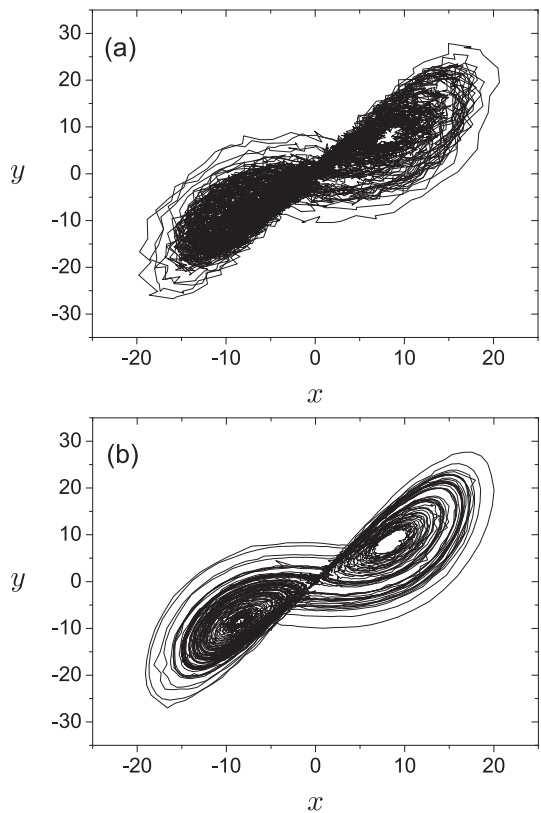}
\caption{(a) Lorenz attractor with 5\% noise added and (b) refined data (100,000 points for each). Relative variation becomes reduced to about 0.05.}
\end{figure}
where $\sigma = 10$, $r=28$ and $b=8/3$. The {\em noisy} orbit $\{X_k\}$ is generated in FIG.\ 1(a) by introducing noise of $\delta \approx 5\%$ of whole system size, which is enough to destroy most important characteristics of the attractor \cite{Kantz}. Our scheme corrects each point $X_k$ into $X'_k$, as depicted in (b), where $20\times20\times20$ neighboring grid points are constructed for each data point and $R_{th}$ is set to be $10$ throughout this calculation. We define {\em relative variance} as
\begin{equation}
\label{relative variance}
e = \frac{\sum_{k=1}^{N}(Y_k - X'_k)^2}{\sum_{k=1}^{N}(Y_k - X_k)^2}
\end{equation}
to quantify the performance of the scheme, where $e < 1$ means that noise is reduced ($e = 0$ for total noise reduction). This demonstration yields $e \approx 0.05$, which implies a high point-to-point correspondence so that this scheme can be categorized as {\em detailed noise reduction} following Ref.\ \cite{Farmer1991}. Similar results are obtained for R\"ossler system. Though Rosa et al. \cite{Rosa} propose that both forward and backward iterations are necessary for high dimensional systems, we do not perform backward one since this noninvertible $M$ lacks time reversal symmetry and thus information flows with only one direction.

So far the full knowledge of dynamics has been assumed for explaining convenience. Although this assumption may be valid in some area, we need to infer dynamics from given raw data in general. Farmer and Sidorowich pointed out that how much noise one can reduce is limited by the accuracy of approximation to the true dynamics \cite{Farmer1991}. At first, we tried to find local linear dynamics as Kotelich and Yorke did \cite{Kost1988}, but it was not quite satisfactory since determining the size of neighborhood was troublesome, that is, too small size often decreases statistical confidence and too large one could not capture the fine structure of the attractor. Looking for alternatives consistent with the above scheme, we noted that the true dynamics would be the most accurate approximation among other candidate models and that our getting closer to the true dynamics could be expressed by longer $m$ in average.

Let us suppose that the parameter $r$ in (\ref{Lorenz}), representing Rayleigh number in convection problem \cite{Strogatz}, is unknown to us. Even if we are given the same data as FIG.\ 1(a), now we should test many Lorenz systems with different $r$ values until finding $r=28$. Figure 2 shows how the choice of $r$ changes the distribution of $\{m_k\}$. We depicted only two cases of $r=28$(correct) and $r=0$(wrong) though we observed the same tendency for intermediate values of $r$. The distribution looks Maxwellian in the vicinity of true dynamics and this Maxwellian region can be reached by processing raw data. We present a qualitative description with a statistical moment of the distribution. Imposing perturbed dynamics, we see that abrasion time goes to zero as our guesses are rejected soon by observations. The average abrasion time $\bar{m} = N^{-1} \sum_{k=1}^N m_k$ rises to 14.71 for $r=28$ while becomes only 5.96 for $r=0$. Let us consider two extreme cases to elucidate basic nature of the distribution : If the underlying dynamics is so trivial (e.g. stable periodic motion) that one can easily discover it, the future orbit is highly predictable and the distribution will be drawn to infinity. As a non-chaotic system contains little information, our noise reduction scheme becomes ineffective with diverging $\bar{m}$. Conversely, if dynamics looks totally unpredictable based on our knowledge, the distribution will collapse to zero point. We again see that noise is not reduced at all, since accuracy of approximation sets an upper bound of reducing performance, as stated above. Thus higher $\bar{m}$ is preferable when dynamics is unknown, while $\bar{m}$ divergence should be avoided when dynamics is known, which may seem contradictory at first. The balance between infinity and zero indicates a status between regularity and randomness, or between perfect predictability and unpredictability. In other words, $\bar{m}$ depends both on the system we observe and the information we have on that system.

\begin{figure}
\includegraphics[angle=0, width=0.4\textwidth]{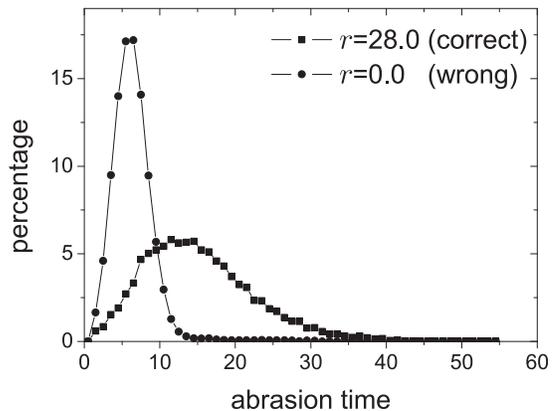}
\caption{The distribution shapes of abrasion time in the vicinity of true dynamics. Deviation from correct $r$ decreases abrasion time in average.}
\end{figure}

From the above arguments, we suggest an algorithm for inferring dynamics : One obtains enough time signals, possibly including noise, and chooses appropriate basis functions specified by a number of parameters. After rough estimation of the parameters, by means of fitting and smoothing algorithms, the higher $\bar{m}$ discovered around those values, the better dynamics inferred. In a brief numerical experiment, we set $(\dot{x}, \dot{y}, \dot{z}) = \vec{M}(x,y,z)$, where components of $\vec{M}$ are second-order polynomials of $x$, $y$, and $z$ with unknown coefficients and we observe that even a crude search can reduce noise with approaching the true dynamics (FIG.\ 3). Tests of 200 random samples around our rough guess give maximum $\bar{m}=5.37$ (only about 37\% of that of true dynamics), but the relative variance $e$ becomes less than 0.7. Advanced parameter searching techniques is expected to yield desirable performance. Such error-tolerance property of {\em $\bar{m}$-method} is supposed to be due to a sort of shadowing effect : a deviated parameter operates as dynamical noise since it is coupled to the dynamics, and an incorrect model can be shadowed by less dynamical-noisy orbits (i.e. with less deviated parameters) within some distances \cite{Hammel1990, Farmer1991}.

\begin{figure}
\includegraphics[angle=0, width=0.4\textwidth]{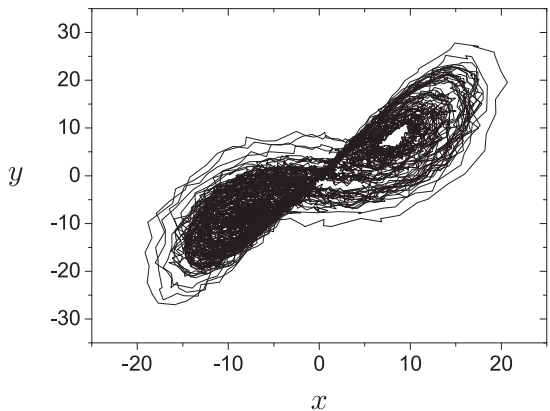}
\caption{100,000 points after a random parameter searching. Compared with FIG.\ 1(a), the orbit becomes a little smoother with relative variance less than 0.7. Advanced searching algorithms are expected to yield better results.}
\end{figure}

In summary, we suggested a nonlinear noise reduction scheme using ideas of information theory, which requires two conditions of chaoticity and well-defined noise level. Since information flow gradually reveals more precise knowledge, it formalizes the problem into rejection of hypotheses instead of minimization. Topological consideration and information-theoretic analysis combined in our scheme provide a concise and easily applicable way for noise reduction. Noise was readily decreased to less than a twentieth for fully known Lorenz system. We introduced abrasion time and proposed its average $\bar{m}$ as a quantifier for inferring dynamics. This $\bar{m}$-method was checked by performing noise reduction, since the accuracy of this inference fundamentally sets a limit on noise reducing capability. It readily yielded the expected noise reduction.

We would like to thank P. -J. Kim, S. -O. Jeong, H. -K. Park, C. Doe, T. -W. Ko, S. Chae and H. -T. Moon for their helpful comments. This work is supported by grant No. R01-1999-000-00019-0 from the Basic Research Program of Korea Science and Engineering Foundation.


\begin{thebibliography}{00}
\bibitem{Kost1988} E. J. Kostelich and J. A. Yorke, Phys. Rev. A {\bf 38}, 1649 (1988)
\bibitem{Kost1989} E. J. Kostelich and J. A. Yorke, Physica {\bf 41D}, 183 (1989)
\bibitem{Hammel1990} S. M. Hammel, Phys. Lett. {\bf 148A}, 421 (1990)
\bibitem{Farmer1991} J. D. Farmer and J. J. Sidorowich, Physica {\bf 47D}, 373 (1991)
\bibitem{Schreiber1993} T. Schreiber, Phy. Rev. E {\bf 47}, 2401 (1993)
\bibitem{Kost1993} E. J. Kostelich and T. Schreiber, Phy. Rev. E {\bf 48}, 1752 (1993)
\bibitem{Davies1994} M. Davies, Physica {\bf 79D}, 174 (1994)
\bibitem{Davies1998} M. Davies, Chaos {\bf 8}, 775 (1998)
\bibitem{Brocker2001} J. Br\"ocker and U. Parlitz, Chaos {\bf 11}, 319 (2001)
\bibitem{Hayes} S. Hayes, C. Grebogi, and E. Ott, Phys. Rev. Lett. {\bf 70}, 3031 (1993)
\bibitem{Cuomo} K. M. Cuomo and A. V. Oppenheim, Phys. Rev. Lett. {\bf 71}, 65 (1993)
\bibitem{Bollt} E. Bollt, Y. -C. Lai, and C. Grebogi, Phys. Rev. Lett. {\bf 79}, 3787 (1997)
\bibitem{OGY} E. Ott, C. Grebogi, and J. A. Yorke, Phys. Rev. Lett. {\bf 64}, 1196 (1990)
\bibitem{Kantz} H. Kantz and T. Schreiber, {\it Nonlinear time series analysis}, Cambridge University Press (1997)
\bibitem{Pecora} L. M. Pecora and T. L. Carroll, Phys. Rev. Lett. {\bf 64}, 821 (1990)
\bibitem{Rosa} E. Rosa, Jr., S. Hayes, and C. Grebogi, Phys. Rev. Lett. {\bf 78}, 1247 (1997)
\bibitem{Andreyev} Y. V. Andreyev, A. S. Dmitriev, E. V. Efremova, and A. N. Anagnostopoulos, Chaos Solitons \& Fractals {\bf 17}, 531 (2003)
\bibitem{Wolf} A. Wolf, J. B. Swift, H. L. Swinney, and J. A. Vastano, Physica {\bf 16D}, 285 (1985)
\bibitem{Shannon} C. E. Shannon, {\it The mathematical theory of communication}, University of Illinois Press (1949)
\bibitem{Brillouin} L. Brillouin, {\it Science and information theory}, Academic Press (1962)
\bibitem{Ott} E. Ott, {\it Chaos in dynamical systems}, Cambridge University Press (1993)
\bibitem{Schittenkopf} C. Schittenkopf and G. Deco, Physica {\bf 94D}, 57 (1996)
\bibitem{Pethel} S. D. Pethel, N. J. Corron, Q. R. Underwood, and K. Myneni, Phys. Rev. Lett. {\bf 90}, 254101 (2003)
\bibitem{Strogatz} S. H. Strogatz, {\it Nonlinear dynamics and chaos}, Addison-Wesley Publishing Company (1994)
\end{thebibliography}
\end{document}